\documentclass[a4paper,11pt]{article}

\usepackage{amssymb}
\usepackage{amsmath}
\usepackage{graphicx}
\usepackage{comment}

\begin{document}

\newcommand{\bin}[2]{\left(\begin{array}{c} \!\!#1\!\! \\  \!\!#2\!\!
\end{array}\right)}
\newcommand{\troisj}[3]{\left(\begin{array}{ccc}#1 & #2 & #3 \\ 0 & 0 & 0
\end{array}\right)}
\newcommand{\sixj}[6]{\left\{\begin{array}{ccc}#1 & #2 & #3 \\ #4 & #5 & #6
\end{array}\right\}}
\newcommand{\neufj}[9]{\left\{\begin{array}{ccc}#1 & #2 & #3 \\ #4 & #5 & #6 \\
#7 & #8 & #9 \end{array}\right\}}

\title{Effect of third- and fourth-order moments on the modeling of Unresolved
Transition Arrays}

\author{J.C. Pain$^{\dag}$\footnote{phone: 00 33 1 69 26 41 85, email: jean-christophe.pain@cea.fr}, F. Gilleron\footnote{CEA/DIF, B.P. 12, 91680 Bruy\`eres-Le-Ch\^atel Cedex, France}, J. Bauche$^{\ddag}$ and C. Bauche-Arnoult\footnote{Laboratoire Aim\'e Cotton, B\^atiment 505, Campus d'Orsay, 91405 Orsay,
France}}

\maketitle


\begin{abstract}
The impact of the third (skewness) and fourth (kurtosis) reduced centered
moments on the statistical modeling of E1 lines in complex atomic spectra is
investigated through the use of Gram-Charlier, Normal Inverse Gaussian and
Generalized Gaussian distributions. It is shown that the modeling of unresolved
transition arrays with non-Gaussian distributions may reveal more detailed
structures, due essentially to the large value of the kurtosis. In the present
work, focus is put essentially on the Generalized Gaussian, the power of the
argument in the exponential being constrained by the kurtosis value. The
relevance of the new statistical line distribution is checked by comparisons
with smoothed detailed line-by-line calculations and through the analysis of
$2p\rightarrow3d$ transitions of recent laser or Z-pinch absorption
measurements. The issue of calculating high-order moments is also discussed
(Racah algebra, Jucys graphical method, semi-empirical approach ...). 
\end{abstract}


\section{\label{int} Introduction}

The detailed calculation of all the electric-dipole (E1) line energies and
radiative strengths in complex atomic spectra is an overwhelming task. In some
circumstances, it can even be useless. Indeed, when the density is sufficiently
high so that the physical broadening mechanisms (Stark, ...) are important
and/or when the number of lines in an energy range becomes large, the lines
coalesce into broad structures. Statistical methods are required because,
experimentally, some quantities cannot be determined individually, but only as
weighted average quantities (this is the case for instance in
emission/absorption
spectra of highly ionized atoms). Moreover, explicit quantum calculations can be
inappropriate, e.g. if the eigenvalues of the hamiltonian are not known with 
sufficient precision. In addition, global methods can reveal physical properties
hidden by a detailed treatment of levels and lines (``one cannot see the wood
for the trees'') \cite{BAUCHE90}.

A transition array \cite{HARRISON31} of E1 lines is characterized by a specific
distribution of photon energy $E$:

\begin{equation}
I(E)=A(E)~\otimes~\Psi(E),
\end{equation}

where the function $A(E)$ assumes that each line is represented by a Dirac
$\delta$-function:

\begin{equation}\label{def:A(E)}
A(E)=C_p~\sum_{a,b}~\frac{N_a}{g_a}~\left(E_{ab}\right)^p~S_{ab}~\delta(E-E_{ab}
)
\end{equation}

which, using the appropriate constant factor $C_p$, represents either the
opacity ($p=1$) or the emissivity of the source ($p=4$). The sum runs over the
upper and lower levels of each line belonging to the transition array. The
density of ions excited in level $a$ is noted $N_a$, and $g_a$ is the degeneracy
of level $a$. The energy of the line $a\rightarrow b$ is 

\begin{equation}
E_{ab}=E_b-E_a=<b|H|b>-<a|H|a>, 
\end{equation}

where $H$ is the Hamiltonian of the system and the line strength $S_{ab}$ is
equal to $|<a|\mathcal{Z}|b>|^2$, where $\mathcal{Z}$ is the $z$ component of
the dipole transition operator. The normalized profile $\Psi(E-E_{ab})$ takes
into account the broadening of the line in the plasma due to natural width,
Doppler effect, ionic Stark effect, electron collisions, etc. In the UTA
(Unresolved Transition Arrays) modeling \cite{BAUCHE79}, the discrete
distribution $A(E)$ can be replaced by a continuous function (usually Gaussian)
which preserves its first- and second-order moments. The density $N_a$ is
assumed to be proportional to its statistical weight $g_a$: $N_a\approx Ng_a/g$,
with $N=\sum_a N_a$ and $g=\sum_a g_a$. In order to avoid the sum over the term
$\left(E_{ab}\right)^p$ in Eq. (\ref{def:A(E)}), the line energy $E_{ab}$ is
replaced by the center of gravity $E_G$ of the transition array, i. e.
$\left(E_{ab}\right)^p\approx \left(E_G\right)^{p}$. These assumptions allow one
to express the moments of this distribution as

\begin{eqnarray}\label{defmu}
\mu_n(A)&=&\frac{\int_{-\infty}^{\infty}A(E)~E^n~dE}{\int_{-\infty}^{\infty}A(E)
~dE}\approx\frac{\sum_{a,b}S_{ab}~\left(E_{ab}\right)^n}{\sum_{a,b}S_{ab}}
\nonumber\\
&\approx &
\frac{\sum_{a,b}[<b|H|b>-<a|H|a>]^n|<a|\mathcal{Z}|b>|^2}{\sum_{a,b}|
<a|\mathcal{Z}|b>|^2}.
\end{eqnarray}

It is possible to derive analytical formulae for the moments $\mu_n(A)$ using
Racah's quantum-mechanical algebra and second-quantization techniques of Judd
\cite{JUDD67}. Such expressions, which depend only on radial integrals, have
been published by Bauche-Arnoult, et al.
\cite{BAUCHE79,BAUCHE82,BAUCHE84,BAUCHE85} for the moments $\mu_n$ (with
$n\leqslant 3$) of several kinds of transition arrays (relativistic or not). It
is useful to introduce the reduced centered moments of the distribution defined
by

\begin{equation}
\alpha_n(A)=\frac{1}{\Omega}\int_{-\infty}^{\infty}
A(E)\left(\frac{E-\mu_1}{\sigma}\right)^ndE,
\end{equation}

where $\mu_1$ is the center-of-gravity of the strength-weighted line energies,
$\sigma=\sqrt{\mu_2-\mu_1^2}$ is the standard deviation and
$\Omega=\int_{-\infty}^{\infty} A(E)~dE$ is the total area of the distribution.
The use of $\alpha_n(A)$ instead of $\mu_n(A)$ allows one to avoid numerical
problems due to the occurence of large numbers. The first values are
$\alpha_0=1$, $\alpha_1=0$ and $\alpha_2=1$. The distribution $A(E)$ is
therefore fully characterized by the values of $\Omega$, $\mu_1$, $\sigma$ and
of the high-order moments $\alpha_n$ with $n>2$. The first four moments are
often sufficient to capture the global shape of the distribution $A(E)$. The
third- and fourth-order reduced centered moments $\alpha_3$ and $\alpha_4$ are
named {\it skewness} and {\it kurtosis}. They quantify respectively the
asymmetry and sharpness of the distribution. The kurtosis is usually compared to
the value $\alpha_4=3$ for a Gaussian. However, the choice of the distribution
does not play any role in the derivation of the moments.


\section{\label{sec2} Impact of the third- and fourth-order moments}

\subsection{\label{subsec21} Gram-Charlier expansion series}

In order to investigate the impact of skewness and kurtosis, it is possible to
use the Gram-Charlier expansion series \cite{KENDALL69}:

\begin{equation} 
GC_n(E)=\frac{\Omega}{\sigma}~\frac{e^{-\frac{u^2}{2}}}{\sqrt{2\pi}}
\left(1+\sum_{k=2}^{n}c_{k}~\text{He}_{k}(u)\right),
\end{equation}

with

\begin{equation}\label{eq:ck}
c_k=\sum_{j=0}^{\text{int}(k/2)}\frac{(-1)^j}{j!(k-2j)!2^j}~\alpha_{k-2j}(A),
\end{equation}

where $u=(E-\mu_1)/\sigma$, $n$ is the order of the moment and
$\text{He}_{k}(u)$ is the Hermite polynomial of order $k$ obeying the recursion
relation:

\begin{equation}
\text{He}_{n+1}(x)=x~\text{He}_{n}(x)-n~\text{He}_{n-1}(x),
\end{equation}

with $\text{He}_0(x)=1$ and $\text{He}_1(x)=x$. The Gram-Charlier expansion
series uses the reduced centered moments $\alpha_k(A)$ of the discrete
distribution $A(E)$. When carried on to infinity, it can be shown that the
Gram-Charlier series is an exact representation of the distribution. Since
$\text{He}_2(x)=x^2-1$, $\text{He}_3(x)=x^3-3x$ and $\text{He}_4(x)=x^4-6x^2+3$,
one finds that the fourth-order Gram-Charlier series reads:

\begin{equation}\label{eqgc4}
GC_4(E)=\frac{\Omega}{\sigma}~\frac{e^{-\frac{u^2}{2}}}{\sqrt{2\pi}}\left[1
-\frac{\alpha_3}{2}\left(u-\frac{u^3}{3}\right)+\frac{(\alpha_4-3)}{24}
(3-6u^2+u^4)\right].
\end{equation}

\begin{figure}
\begin{center}
\includegraphics[width=10cm]{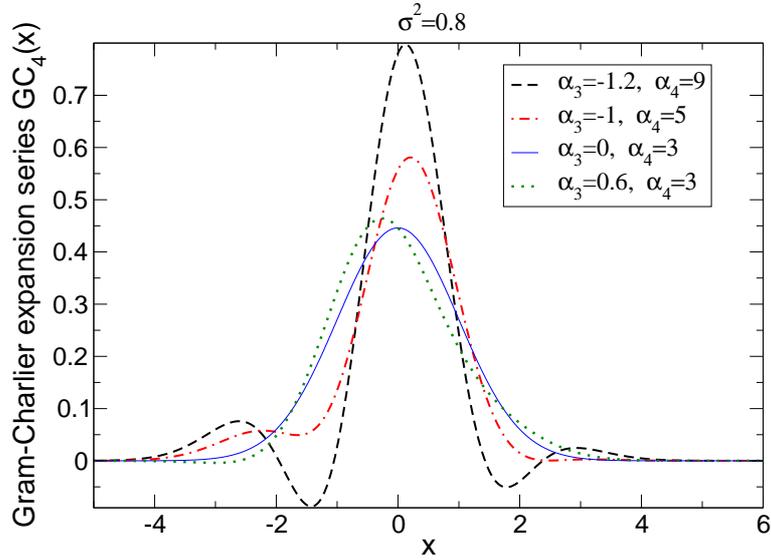}
\end{center}
\caption{(Color online) Fourth-order symmetrical Gram-Charlier GC$_4$
distributions for several
values
of the coefficients $\alpha_3$ and $\alpha_4$. The full line corresponds to the
Gaussian
function.}\label{fig1}
\end{figure}

One of the main disadvantages of the Gram-Charlier profile is that in certain
circumstances (see Fig. \ref{fig1}), it exhibits some negative features. 

\begin{figure}
\begin{center}
\includegraphics[width=10cm]{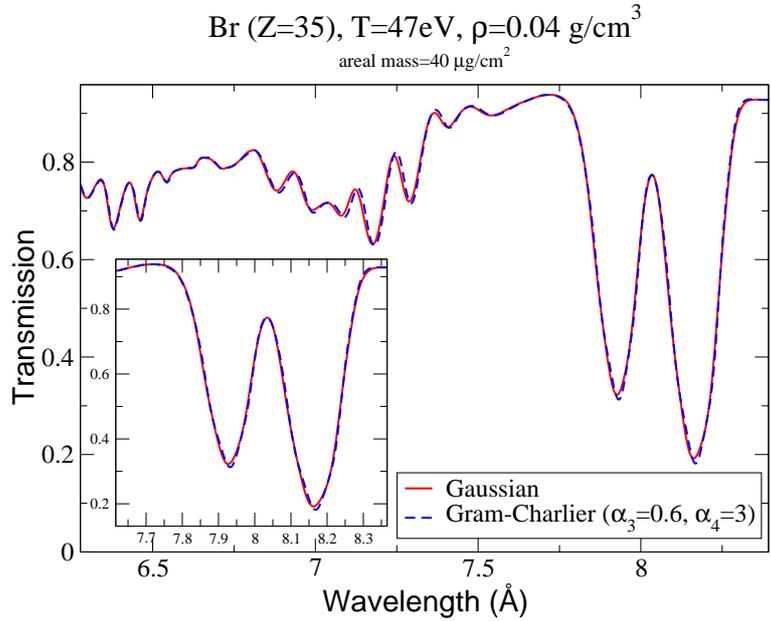}
\end{center}
\caption{(Color online) Impact of a skewed Gaussian for modeling the opacity
lines of a bromine
plasma ($T$=47 eV and $\rho$=0.04 g/cm$^3$) in the $2p\rightarrow 3d$ range.}
\label{fig2}
\end{figure}

The levels of an electronic configuration verify approximate symmetries, which
are called couplings in atomic spectroscopy. In fact, highly asymmetrical line
distributions can be found due to some large $G^1$ exchange Slater integral.
This may happen in arrays of the type $l^{N+1}-l^Nl'$, which are very numerous,
because in highly charged ions, the transitions decaying to the ground
configuration belong to arrays of this type. The skewness increases steeply as a
function of atomic number $Z$ along an iso-electronic sequence. Assuming
hydrogenic behaviour, Slater and spin-orbit integrals vary as $Z$ and $Z^4$,
respectively. It was shown in that case by Bauche et al. \cite{BAUCHE88} that
$\alpha_3$ behaves roughly as $Z^6$. Therefore, asymmetry is particularly
pronounced for highly ionized heavy atoms. They can also occur in situations
where the spin-orbit interactions are strong enough to split the transition
array (medium- or high-$Z$ elements). In the latter case, the asymmetrical shape
of the array can also be restored by considering the superposition of
symmetrical subarrays. In fact, the impact of skewness is usually small in the
conditions typical of the laser experiments conducted to date (density of the
order of 0.01
g/cm$^3$ and temperature of a few tens of eV). Fig. \ref{fig2} displays the
comparison between a skewed Gaussian (fourth-order Gram-Charlier (see Eq.
(\ref{eqgc4})) with $\alpha_3=0.6$ and $\alpha_4=3$ compared to a Gaussian
modeling of lines in the case of a bromine plasma at $T$=47 eV and $\rho$=0.04
g/cm$^3$. The spectra are almost indistinguishable.

\subsection{\label{subsec22} Normal Inverse Gaussian}

\begin{figure}
\begin{center}
\includegraphics[width=10cm]{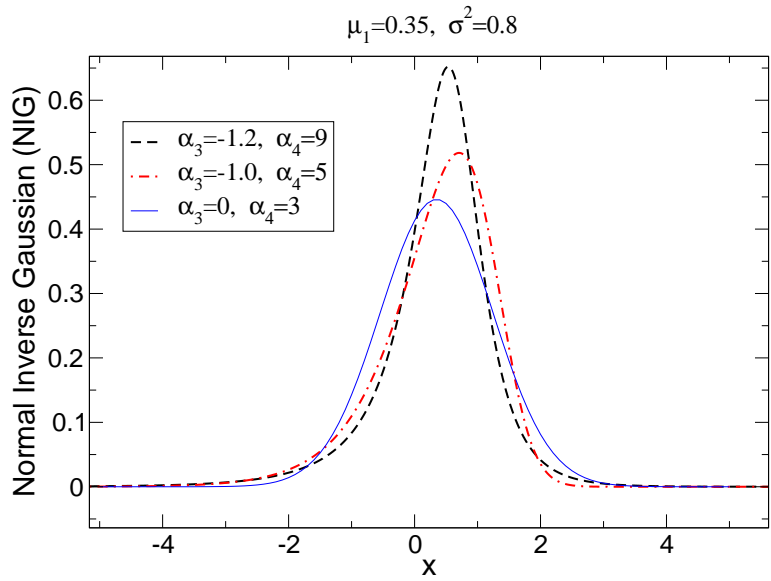}
\end{center}
\caption{(Color online) Examples of NIG distributions corresponding to different
values of the
parameters $\alpha$, $\beta$, $\delta$ and $\mu$. The full line corresponds to
the Gaussian function.}
\label{fig3}
\end{figure}

\begin{figure}
\begin{center}
\includegraphics[width=10cm]{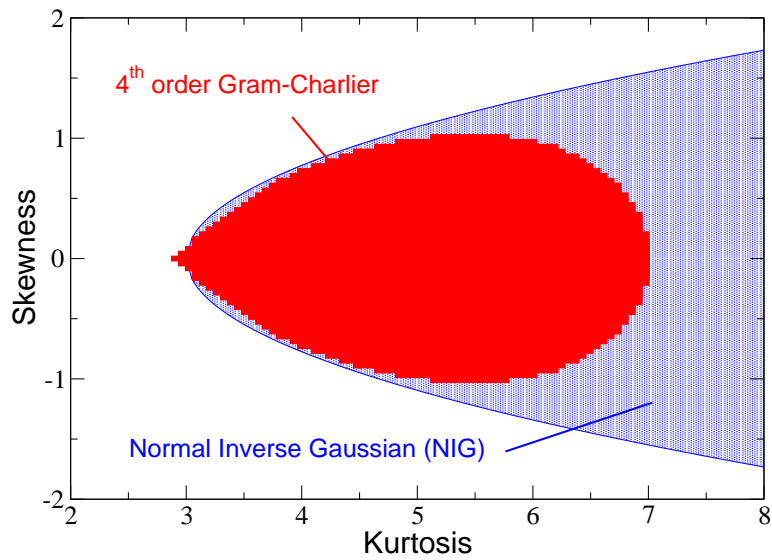}
\end{center}
\caption{(Color online) Validity domain of the NIG distribution (defined by
$|\alpha_3|<\sqrt{\frac{3}{5}(\alpha_4-3)}$) and positivity domain of the GC$_4$
distribution.}
\label{fig4}
\end{figure}

If one wants to take into account the effect of asymmetry, it can be fruitfully
done using the NIG (Normal Inverse Gaussian) distribution \cite{BARNDORFF97}:

\begin{equation}
\text{NIG}(E)=\frac{\delta\alpha
e^{\delta\sqrt{\alpha^2-\beta^2}+\beta(E-\mu)}}{\delta^2+(E-\mu)^2}K_1(\alpha
\sqrt{\delta^2+(E-\mu)^2}),
\end{equation}

where $K_1$ is a modified Bessel function of the third kind. The four parameters
$\alpha$, $\beta$, $\delta$ and $\mu$ are obtained directly from the knowledge
of $\mu_1$, $\sigma$, $\alpha_3$ and $\alpha_4$. The corresponding relations as
well as the role of the latter parameters can be found in Table
\ref{param_fig3}. Unlike the Gram-Charlier expansion series, the NIG
(see Fig. \ref{fig3}) cannot have negative values. Moreover, the validity domain
of the NIG is wider than the positivity domain of Gram-Charlier distribution, as
shown in Fig. \ref{fig4}. The NIG distribution enables one to model symmetric or
asymmetric distributions with possibly long tails in both directions. The tails
are much more prominent than in the Gaussian distribution.

\subsection{\label{subsec23} Generalized Gaussian}

An interesting choice to study solely the effects of the kurtosis is the
generalized Gaussian distribution (GG), defined by:

\begin{equation}\label{eq:gg}
\text{GG}(E)=\frac{\Omega\nu}{\sigma}~\frac{e^{-\left|\frac{u}{\lambda}
\right|^\nu}}{2\lambda~\Gamma(\frac{1}{\nu})}
\quad\text{with}\quad
\lambda=\sqrt{\frac{\Gamma(\frac{1}{\nu})}{\Gamma(\frac{3}{\nu})}},
\end{equation}

where $\nu$ is a positive real number, and $\Gamma(x)$ is the ordinary gamma
function. The even-order moments of a GG function read:

\begin{equation}\label{eq:momgg}
\alpha_{2k}(\text{GG})=\lambda^{2k}~\frac{\Gamma(\frac{1+2k}{\nu})}
{\Gamma(\frac{1}{\nu})},
\end{equation}

whereas the odd-order moments are null, $\alpha_{2k+1}(P)=0$, since the GG is
symmetric. The parameter $\nu$ can be obtained by constraining the kurtosis
coefficient, and thus solving the equation

\begin{equation}\label{rootf}
\alpha_4=\frac{\Gamma(1/\nu)\Gamma(5/\nu)}{\Gamma(3/\nu)^2}.
 \end{equation}
 
\begin{figure}
\begin{center}
\includegraphics[width=10cm]{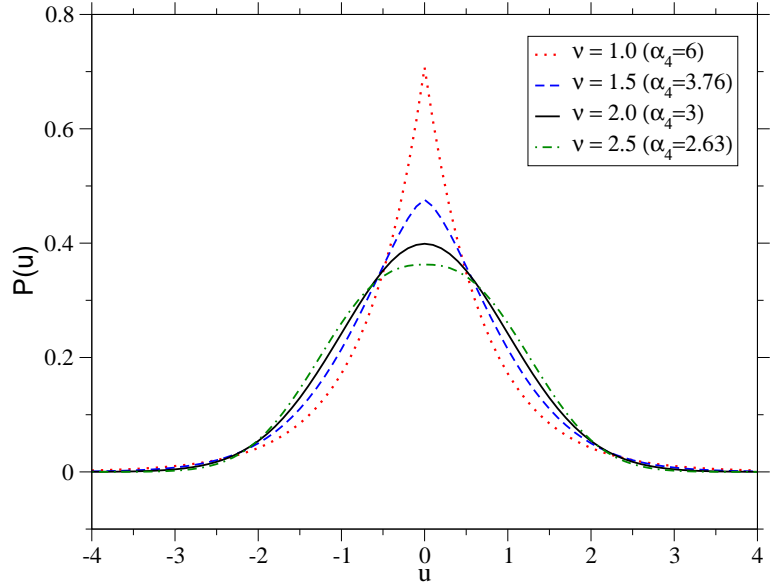}
\end{center}
\caption{(Color online) Examples of GG distributions (corresponding to different
values of the
exponent $\nu$).}
\label{fig5}
\end{figure}

This distribution is represented in Fig. \ref{fig5} for several values of the
parameter $\nu$ (i.e. for different values of the kurtosis $\alpha_4$). The GG
function has interesting properties. It is a simple increasing (decreasing)
function for $u<0$ ($u>0$), without negative values in contrast with GC series.
The Gaussian ($\nu=2$) and the Laplace ($\nu=1$) distributions are special cases
of GG functions with a kurtosis coefficient equal to $3$ and $6$, respectively.
The discontinuity of the derivative at $u=0$ for $0<\nu\leqslant 1$ disappears
with the convolution by another function. The full width at half maximum (FWHM)
of a GG function is $2~\sigma\lambda~(\log 2)^\frac{1}{\nu}$, and therefore
depends on the distribution itself. For example, the above formula gives
$\text{FWHM}=2.35~\sigma$ for a Gaussian ($\nu=2$) and $\text{FWHM}=
0.98~\sigma$ for a Laplace distribution ($\nu=1$). The root of Eq. (\ref{rootf})
is fairly well approximated by the fitting function

\begin{equation}
\nu=1.62796\left[\ln(\alpha_4-0.783143)\right]^{\frac{0.796349}{\alpha_4^2}-1}.
\end{equation}

\subsection{\label{subsec24} Comparisons of the distributions}

\begin{figure}
\begin{center}
\includegraphics[width=10cm]{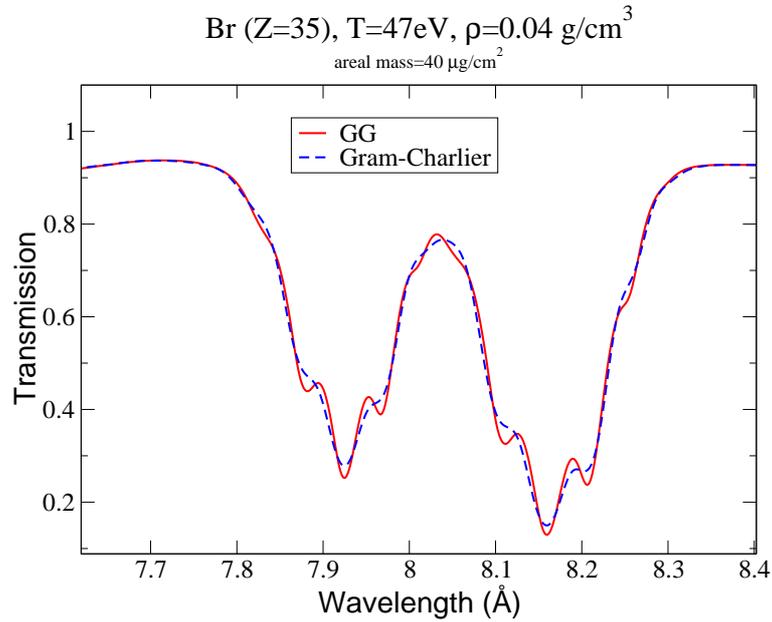}
\end{center}
\caption{(Color online) Comparison of a transmission spectrum (same as Fig.
\ref{fig2}) with a
GG$_{\nu=1}$ profile and a symmetrical GC$_4$ profile both characterized by the
same first four moments (fixing in particular $\alpha_3$=0 and $\alpha_4$=6).}
\label{fig6}
\end{figure}

Fig. \ref{fig6} displays the comparison of a transmission spectrum (for the same
case as Fig. \ref{fig2}) with GG and GC profiles with $\alpha_3$=0 and
$\alpha_4$=6. We observe that both distributions give rise to slightly different
spectra, although they are characterized by the same first four moments. The
differences are the signature of high-order moments $\alpha_n$ ($n>4$). Indeed,
the first constained moments fix ipso facto the higher-order unconstrained
moments. This can be seen in Table \ref{highermoments}, for the transition array
$3d^6-3d^54p$ in Br XII, which shows the values of $\alpha_k$ up to the order
$14$ for several distributions. The exact values of the moments are calculated
with Cowan's atomic structure code \cite{COWAN81} (routines RCN/RCN2/RCG). They
are compared with the values for the GG function, using Eq. (\ref{eq:momgg}),
with the values for the fourth-order GC series (obtained by setting $c_k=0$ for
$k\ge 6$) and for the NIG function (requiring $\alpha=0.82$, $\beta=0$,
$\delta=3.17$ and $\mu=154.18$). Functions GG, GC$_4$ and NIG are constrained by
the moments of order $k\le 4$ of the exact distribution. Values for the Gaussian
function are also shown. It is observed, for this particular case, that the GG
function is slightly closer to the exact values than the other distributions. 

\begin{figure}
\begin{center}
\includegraphics[width=10cm]{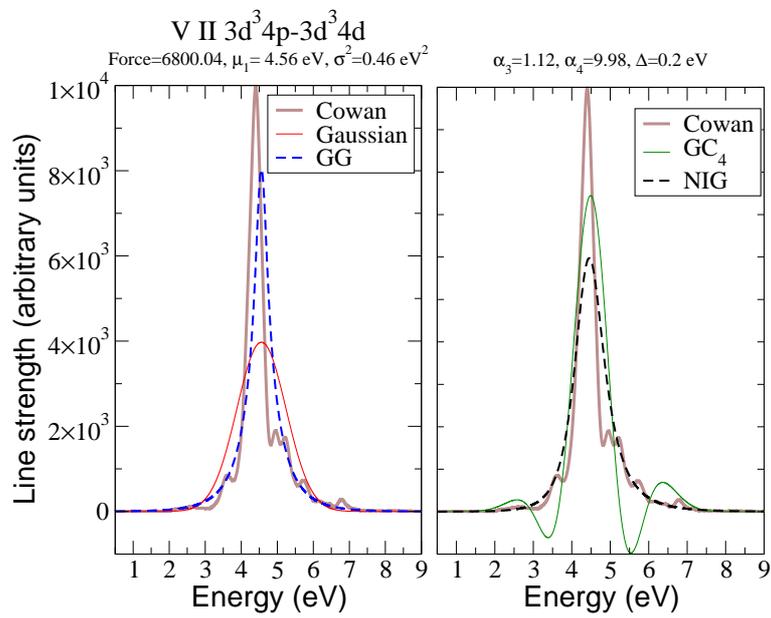}
\end{center}
\caption{(Color online) Comparison of the line distribution of transition array
$3d^34p\rightarrow 3d^34d$ for V II calculated with Cowan's code, and modeled by
a Gaussian, GG, GC$_4$ and NIG profile. All the distributions are convolved by 
a Gaussian having a FWHM=0.2 eV.}
\label{fig7}
\end{figure}

\begin{figure}
\begin{center}
\includegraphics[width=10cm]{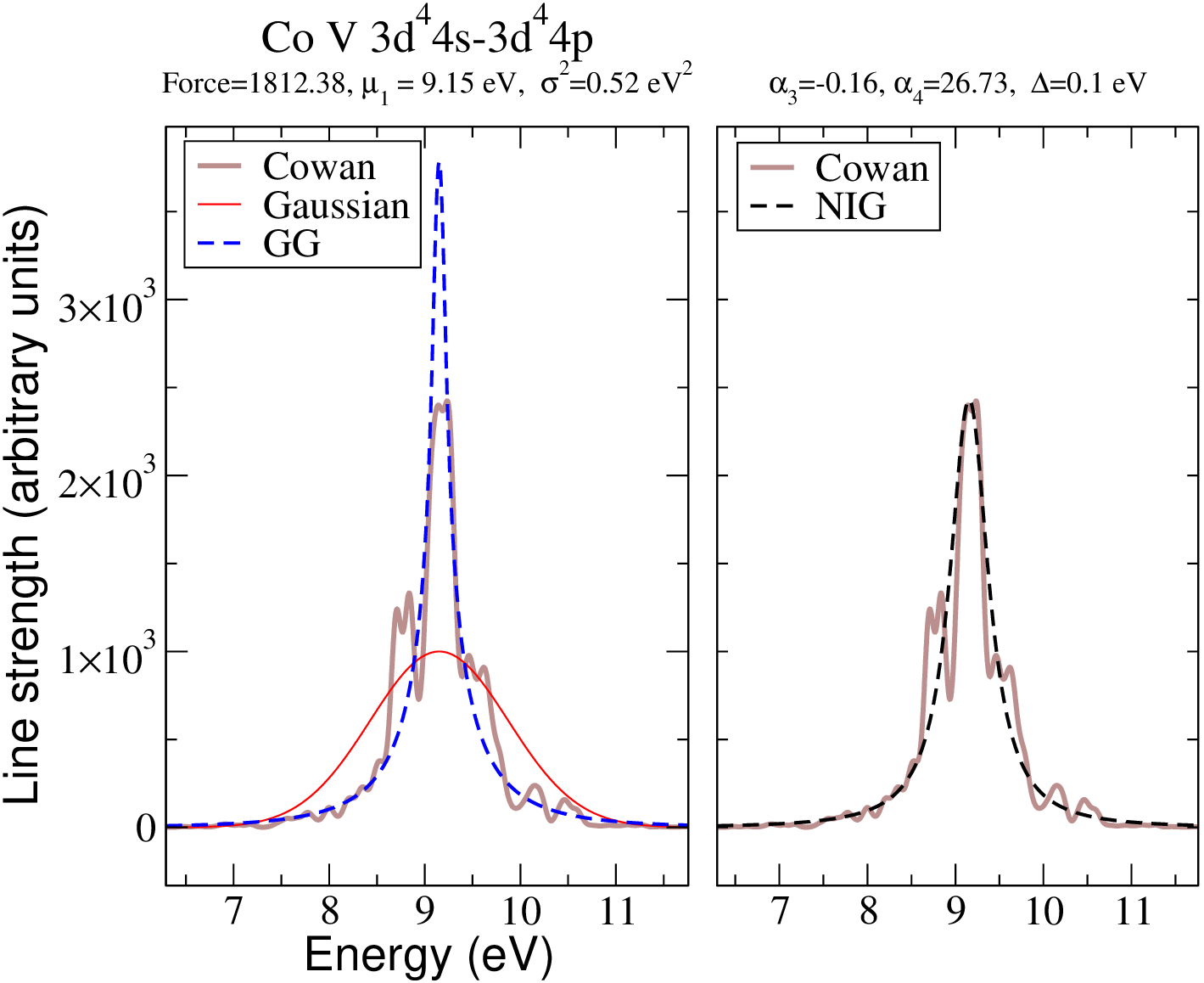}
\end{center}
\caption{(Color online) Same as Fig. \ref{fig7} for the transition array
$3d^44s\rightarrow
3d^44p$ of Co V with a Gaussian linewidth of FWHM=0.1 eV.}
\label{fig8}
\end{figure}

Transition arrays of the kind $l^Nl'-l^Nl''$ are known to be much sharper than a
Gaussian, because of the strong selection rules on the core $l^N$. Fig.
\ref{fig7} displays a comparison of the line distribution of transition array
$3d^34p\rightarrow 3d^34d$ for V II calculated with Cowan's code, and modeled by
Gaussian, GG, GC$_4$ and NIG profiles. All the distributions are convolved by a
Gaussian having a FWHM=0.2 eV. We can see that the width and heigth of the
distribution are not well depicted by the Gaussian. The Gram-Charlier GC$_4$
distribution has negative features. The GG and NIG distributions are more suited
in that case. Fig. \ref{fig8} displays the same kind of comparisons for the
transition array $3d^44s\rightarrow 3d^44p$ of Co V with a Gaussian of FWHM=0.1
eV. The Gram-Charlier GC$_4$ has not been represented since it exhibits huge
non-physical bumps. In that case also the GG and NIG distributions give a better
agreement with the exact distribution than the Gaussian. This visual agreement
has been confirmed by RMS calculations. Finally, Fig. \ref{fig9} displays a
comparison of the profiles for the transition array $3d^34p\rightarrow 3d^34d$
for Co VI, the distributions being convolved by a Gaussian having a FWHM=0.3 eV.
In that case the distribution is asymmetric, and therefore the NIG appears to be
the best distribution, but the GG still gives better results that the
Gaussian. 

We may conclude that the Gaussian is not a good representation for most
transition arrays and that GG or NIG are usually better choices than GC when
accounting for skewness and/or kurtosis coefficients.

\section{\label{sec3} Interpretation of recent $2p\rightarrow3d$ absorption
experiments}

Chenais-Popovics et al. \cite{CHENAIS00} measured the absorption of the
$2p\rightarrow 3d$ transitions of iron in the range 16.4-17.2 \AA. The sample,
heated by the thermal radiation of a gold spherical hohlraum, was irradiated by
the laser ASTERIX IV. The plasma is assumed to be in local thermodynamic
equilibrium (LTE) at a temperature $T=20$ eV and a density $\rho=0.004$
g/cm$^3$. An interpretation using a Detailed Configuration Accounting (DCA)
calculation based on SCO code \cite{BLENSKI00} is presented in Fig. \ref{fig10}.
It can be seen that the departure from the Gaussian allows one to better
reproduce the depth of the successive shoulders in the spectrum which correspond
to the transitions
$2p_{1/2}\rightarrow 3d_{3/2}$, $2p_{3/2}\rightarrow 3d_{3/2}$ and
$2p_{3/2}\rightarrow 3d_{5/2}$ of several ions.

The Z-pinch at Sandia National Laboratory was used by Bailey et al.
\cite{BAILEY03} to measure absorption of
NaBr samples. The main purpose of the experiment was to study the
$2p\rightarrow 3d$ transitions in bromine ionized into the M-shell. Electron
temperature
and density, obtained from the analysis of the sodium lines (which are separated
from the bromine lines), are respectively
$50(\pm 4)$ eV and $3 (\pm 1) 10^{21}$ cm$^{-3}$. The spectral resolution is
about $1.5$ eV. The spin-orbit interactions clearly
separate the $2p_{1/2}\rightarrow 3d_{3/2}$ and $2p_{3/2}\rightarrow 3d_{5/2}$
structures. Fig. \ref{fig11} displays the experimental spectrum and the
calculated DCA spectra at $T=47$ eV obtained with a Gaussian (dashed curve) or a
GG$_{\nu=1}$ function (full line) for the shape of the transition arrays. Some
details of the $2p\rightarrow 3d$ structures, hidden in the Gaussian
description, appear
clearly in the new modeling and give a much better agreement with the
experiment.
This provides not only a better identification of the experimental features, but
also a possible refinement of the temperature and density diagnostic.


\section{\label{sec4} The evaluation of high-order moments: a challenging task}

\subsection{\label{subsec41} Complexity of the calculation}

Moments $\mu_n$ defined in Eq. (\ref{defmu}) can be expressed in terms of sums
involving products of radial integrals. A first difficulty is that the number of
terms in that sum increases very rapidly as the order increases. For instance,
assuming that the transition array of interest is characterized by $q$ different
Slater integrals ($R$) and $r$ different spin-orbit integrals ($\zeta$), the
number of terms of the form $\underbrace{R\cdots R}_{\text{n
terms}}\underbrace{\zeta\cdots\zeta}_{\text{p terms}}$ is

\begin{equation}
N_{\text{max}}=S_q(n)\times S_r(p),
\end{equation}

where 

\begin{equation}
S_t(v)=\bin{v+t-1}{t-1}=\frac{(v+t-1)!}{(t-1)!\;v!}.
\end{equation}

Table \ref{momnumbers} gives $N_{\text{max}}$ for each kind of product of such
integrals and for each of the first four moments $\mu_n$, $n$=1, 2, 3 and 4.
Obviously, this is a maximum since some of the terms do not exist due to
angular-momentum symmetries. In particular, terms of the kind $R\cdots R\zeta$
containing only one $\zeta$ integral are not included since their contribution
is zero due to the Land\'e center-of-gravity rule \cite{BAUCHE79,COWAN81}. One
can see that the number of terms increases very fast with respect to the order
of the moment. The second difficulty is that the full derivation of each term of
the sum is very cumbersome, due to the complex angular-momentum algebra. Uylings
has shown \cite{UYLINGS84} that the trace of a $k-$electron operator in the
space of the states of an electronic configuration $l^N$ is proportional to
$\bin{4l+2-k}{N-k}$. Using second-quantization techniques of Judd \cite{JUDD67},
one finds that the fourth-order moment of $l^{N+1}\rightarrow l^Nl'$ involves 1-
to 10-electron operators and reads therefore

\begin{equation}
\mu_4[l^{N+1}\rightarrow l^Nl']=\sum_{k=1}^{10}\bin{4l+2-k}{N+1-k}p_k,
\end{equation}

where coefficients $p_k$ have to be determined from special cases and symmetry
properties (such as complementarity or anti-complementarity). Particular cases
are difficult to calculate when dealing with more than two electrons
(coefficients of fractional parentage are then required to account for the Pauli
exclusion principle). But the calculation is a hard task even with only two
electrons ; for instance, considering the simple transition array
$l^2\rightarrow ll'$ (corresponding to $N$=1), one has:

\begin{equation}
\mu_4[l^2\rightarrow
ll']=\sum_{k_1,k_2,k_3,k_4}A_{k_1,k_2,k_3,k_4}F^{k_1}(ll)F^{k_2}(ll)F^{k_3}(ll)
F^{k_4}(ll)+\cdots,
\end{equation}

with

\begin{equation}
A_{k_1,k_2,k_3,k_4}\propto
(2-(-1)^L)[L]\sixj{l}{l}{k_1}{l}{l}{L}\sixj{l}{l}{k_2}{l}{l}{L}
\sixj{l}{l}{k_3}{l}{l}{L}\sixj{l}{l}{k_4}{l}{l}{L},
\end{equation}

where $[L]=2L+1$ and $F^{k_i}(ll)$ is a direct Slater integral describing
electrostatic interactions between two electrons inside orbital $l$. The
quantity $A_{k_1,k_2,k_3,k_4}$ contains also the $L$-independent product 

\begin{equation}
[<l||C^{(k_1)}||l><l||C^{(k_2)}||l><l||C^{(k_3)}||l><l||C^{(k_4)}||l>]^2,
\end{equation}

where

 \begin{equation}
<l||C^{(k)}||l>=(-1)^l[l]\troisj{l}{k}{l},
\end{equation}
 
is the reduced matrix element of the spherical function operator $C^{(k)}$, with
the phase convention of Ref. \cite{COWAN81}. Using the graphical representation
of 6-$j$ symbols \cite{JUCYS64}, a compact graph is obtained for the product of
four 6-$j$ symbols involved in $A_{k_1,k_2,k_3,k_4}$ (see Fig. \ref{fig12}). The
summation over $L$ leads to a 12-$j$ symbol of the second kind, and the
summation over $L$ with the phase factor $(-1)^L$ to a 12-$j$ symbol of the
first kind \cite{VARSHALOVICH88}. Therefore, the ``direct'' calculation of the
moments is complicated and the final result would take many pages. However, some
algorithms have been proposed in order to evaluate these high-order moments.
Karazija et al. \cite{KARAZIJA91a, KARAZIJA91b, KARAZIJA95} expressed the
spectral moments by averages of the products of operators and formulated a
general group-diagrammatic method for the evaluation of their explicit
expressions. Oreg et al. \cite{OREG90} used the fact that the moments reduce to
configuration averages of n-boby symmetrical operators (nBSTOs). For that
purpose, they introduced the concept of an n-electron minimal configuration,
relative to the actual (N-electron) configuration average. Their algorithm uses
graphical technique (routine NJGRAF \cite{BAR88}) in order to derive the
dependence of the averages on the orbital quantum numbers in terms of closed
diagrams. 

\subsection{\label{subsec42} Estimation of the kurtosis coefficient}

Let us consider the following ideal distribution of line energies and amplitudes
\cite{BAUCHE91,GILLERON07}:

\begin{equation}\label{ideal}
D(\epsilon,a)=\frac{L}{\sqrt{2\pi
v}}\exp\left[-\frac{\epsilon^2}{2v}\right]\frac{\lambda}{2}\exp[-\lambda|a|],
\end{equation}

where $v$ is the unweighted variance of the line energies:

\begin{equation}\label{vnonpond}
v=\frac{1}{L}\sum_{a,b}\left(E_{ab}\right)^2-\frac{1}{L^2}
\left[\sum_{a,b}E_{ab}\right]^2,
\end{equation}

$\epsilon$ is the line energy, $a$ the line amplitude and $L$ the number of
lines. The average value of any quantity $q(\epsilon,a)$ is given by

\begin{equation}
<q>=\frac{1}{L}\int\int d\epsilon ~da ~D(\epsilon,a)~q(\epsilon,a).
\end{equation}

Following Ref. \cite{BAUCHE91}, the correlation law between the line energies
and amplitudes is taken to be

\begin{equation}
v_a(\epsilon)=\frac{2}{\lambda^2}=\gamma\exp[-\beta|\epsilon|],
\end{equation}

where $v_a$ is the variance of the line amplitudes, corresponding to the average
line strength between two levels. The parameters $\gamma$ and $\beta$ are
determined by requiring the conservation of the total strength
$S=\sum_{a,b}S_{ab}=L<a^2>$ and of the weighted variance $v_w$ of the line
energies:
 
\begin{equation}\label{vpond}
v_w=\frac{1}{S}\sum_{a,b}S_{ab}\left(E_{ab}\right)^2-\frac{1}{S^2}
\left[\sum_{a,b}E_{ab}\right]^2=\frac{L}{S}<a^2\times\epsilon^2>.
\end{equation}

The kurtosis is given by

\begin{equation}
\alpha_4=\frac{S}{L}\frac{<a^2\times\epsilon^4>}{<a^2\times\epsilon^2>^2},
\end{equation}

which leads to the following expression:

\begin{equation}\label{kurgp}
\alpha_4=\frac{1}{\omega^2}[-2+(5+X^2)\omega], 
\end{equation}

where $\omega=v_w/v$, and $X$ is the root of the following equation:

\begin{equation}
(1+X^2-\omega)\exp\left[\frac{X^2}{2}\right]\text{erfc}\left[\frac{X}{\sqrt{2}}
\right]-\sqrt{\frac{2}{\pi}}X=0.
\end{equation}

The energy-amplitude correlation is such that, considering the case where the
spin-orbit interaction is weak, the stronger lines are found closer to the
center of gravity than the weaker ones. Due to these correlations (propensity
rule), the transition array is expected to be sharp. In other words, the
variance of the line energies is always smaller when it is calculated with a
weight equal to the line strength (see Eq. (\ref{vpond})) than when it is not
(see Eq. (\ref{vnonpond})). Eq. (\ref{ideal}) implies that the energy-dependent
distribution of line strengths $S(\epsilon)$ (i.e. the shape of the transition
array) can be written

\begin{equation}
S(\epsilon)=v_a(\epsilon)\frac{L}{\sqrt{2\pi
v}}\exp\left[-\frac{\epsilon^2}{2v}\right],
\end{equation}

which means that, if Eq. (\ref{ideal}) is fulfilled, the resulting shape has
necessarily a kurtosis between 3 (Gaussian) and 6 (Laplace). Table
\ref{kurtoapprox} shows the values of $\alpha_4$ estimated from Eq.
(\ref{kurgp}) compared to the exact ones (i.e. calculated from Cowan's code). 


\section{Conclusion}

The UTA formalism gives the first two moments of the distribution of
electric-dipolar lines. It does not contain any assumption as concerns the
modeling function. Usually, the chosen distribution is the Gaussian (kurtosis
$\alpha_4$=3), but detailed calculations using Cowan's code show that it is
often not the most proper distribution, since it does not account for the
skewness and kurtosis coefficients. In this article, several other distributions
have been studied in an attempt to include the effects of high-order moments.
Gram-Charlier expansion series, which consists in a Gaussian multiplied by a
linear combination of Hermite polynomials, can have negative values. The Normal
Inverse Gaussian function is well suited in order to account for asymmetry and
flatness. However, the impact of asymmetry is usually small in the conditions
typical of usual photo-absorption experiments, and the Generalized Gaussian
function seems to be a good choice in order to account for the specific effect
of the kurtosis \cite{GILLERON08}. With such a modeling, more detailed
structures appear in spectra, leading to a better agreement with recent
$2p\rightarrow3d$ absorption experiments and improving in that way the
temperature and density diagnostics. The analytical calculation of high-order
($n\ge 3$) moments is possible (see the works of Karazija et al.
\cite{KARAZIJA91a,KARAZIJA91b,KARAZIJA95} and Oreg et al. \cite{OREG90}), but
difficult. An approximate method, relying on a physically realistic distribution
of line amplitudes and energies, has been proposed in order to estimate the
kurtosis. The next step will be the calculation of high-order STA (Super
Transition Arrays \cite{BAR89}) moments. 

\vspace{1cm}

\textbf{Acknowledgements}

The authors would like to thank J.E. Bailey, and C. Chenais-Popovics for
providing the experimental spectra.

\clearpage

\begin{table}
\begin{center}
\begin{tabular}{|c|c|c|}\hline
Parameter & Expression & Role \\ \hline \hline
$\alpha$  &
$\frac{3\sqrt{3\tilde{\alpha}_4-4\alpha_3^2}}{\sigma(3\tilde{\alpha}_4
-5\alpha_3^2)}$  & Tail heaviness   \\ \hline
$\beta$   & $\frac{3\alpha_3}{\sigma(3\tilde{\alpha}_4-5\alpha_3^2)}$           
          & Asymmetry   \\ \hline
$\delta$  & $\frac{3\sigma\sqrt{3\tilde{\alpha}_4-5\alpha_3^2}}{3
\tilde{\alpha}_4-4\alpha_3^2}$    
          & Scale parameter  \\ \hline
$\mu$     & $\mu_1-\frac{\delta\alpha_3}{\sqrt{3\tilde{\alpha}_4-5\alpha_3^2}}$ 
          & Location  \\ \hline \hline
\end{tabular}
\caption{Parameters of the NIG distribution. The quantity
$\tilde{\alpha}_4=\alpha_4$-3 represents the excess kurtosis.}\label{param_fig3}
\end{center}
\end{table}

\vspace{1cm}

\begin{table}
\begin{center}
\begin{tabular}{|c|c|c|c|c|c|}\hline
  order &  exact      & GG$_{\nu=1.356}$ & GC$_4$   & NIG       & Gaussian  \\
\hline \hline
    4 & 4.15      & 4.15     & 4.15    & 4.15      & 3       \\ \hline
    6 & 33.59     & 34.73    & 32.22   & 38.81     & 15      \\ \hline
    8 & 433.98    & 466.46   & 346.10  & 665.06    & 105     \\ \hline
   10 & 7772      & 8954     & 4562    & 18480     & 945     \\ \hline
   12 & 1.73 $10^5$ & 2.29 $10^5$ & 0.70 $10^5$ & 7.66 $10^6$ & 10395   \\
\hline
   14 & 4.52 $10^6$ & 7.45 $10^6$ & 1.22 $10^6$ & 4.46 $10^7$ & 135140  \\
\hline \hline
\end{tabular}
\caption{Values of the even reduced centered moments $\alpha_k$ of several
distributions representing the transition array $3d^6\rightarrow 3d^54p$ in Br
XII. Exact:
calculation with Cowan's code; GG: generalized Gaussian function with
$\nu$=1.356; GC$_4$: fourth-order Gram-Charlier series with $\alpha_3$=0 and
$\alpha_4$=4.15; NIG: Normal Inverse Gaussian with $\alpha$=0.82, $\beta$=0,
$\delta$=3.17, $\mu$=154.18.}\label{highermoments}
\end{center}
\end{table}

\vspace{1cm}

\begin{table}
\begin{center}
\begin{tabular}{|c|c|c|c|c|}\hline
Order     & 1 & 2 & 3 & 4 \\ \hline \hline
Products involved  & $R$ & $RR$, $\zeta\zeta$ & $RRR$, $R\zeta\zeta$,
$\zeta\zeta\zeta$ & $RRRR$, $RR\zeta\zeta$, $R\zeta\zeta\zeta$,
$\zeta\zeta\zeta\zeta$ \\ \hline
Max. number of terms & 4 & 16 & 54 & 150 \\ \hline \hline
\end{tabular}
\caption{Number of terms involving products of radial integrals in the first
moments $\mu_n$ of transition array $l^{N+1}\rightarrow
l^Nl'$.}\label{momnumbers}
\end{center}
\end{table}

\vspace{1cm}

\begin{table}
\begin{center}
\begin{tabular}{|c|c|c|}\hline
Transition array & \multicolumn{2}{c|}{kurtosis}\\ \cline{2-3}
 & exact & formula \\\hline \hline
Fe V $3d^4\rightarrow 3d^34p$    & $4.1$  & $3.8$   \\ \hline
Pd VII $4d^4\rightarrow 4d^35p$  & $4.8$  & $4.5$   \\ \hline
Sn XXIX $3d^4\rightarrow 3d^34p$ & $5.9$  & $5.1$  \\ \hline \hline
\end{tabular}
\caption{Estimation of the kurtosis from Eq. (\ref{kurgp}).}\label{kurtoapprox}
\end{center}
\end{table}

\clearpage

\begin{figure}
\begin{center}
\includegraphics[width=10cm]{fig9.eps}
\end{center}
\caption{(Color online) Same as Fig. \ref{fig7} for the transition array
$3d^34p\rightarrow
3d^34d$ of Co VI with a Gaussian linewidth of FWHM=0.3 eV.}
\label{fig9}
\end{figure}

\begin{figure}
\begin{center}
\includegraphics[width=10cm]{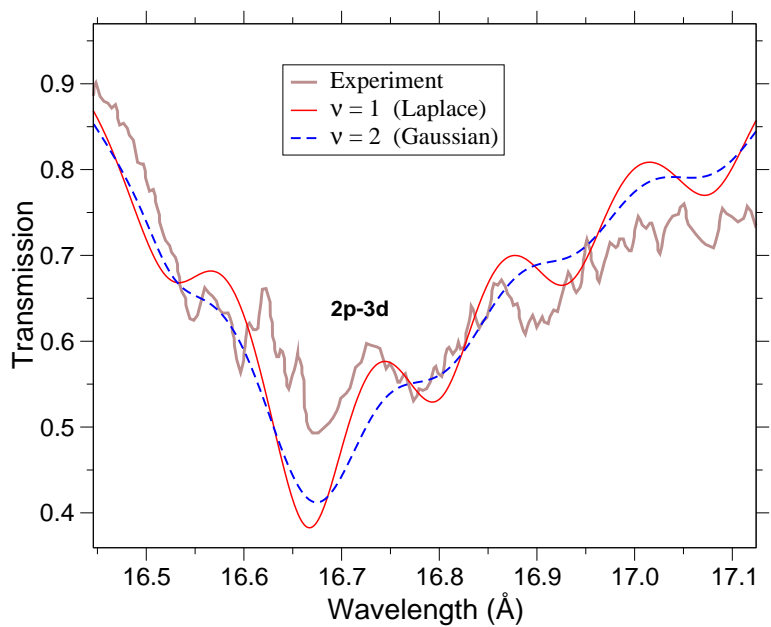}
\end{center}
\caption{(Color online) Absorption spectrum of iron measured by Chenais-Popovics
et al. \cite{CHENAIS00}. The DCA calculations are performed at $T$=20 eV and
$\rho$=0.004 g/cm$^3$ assuming either a Gaussian (dashed line) or a GG$_{\nu=1}$
(full line) profile for the statistical UTA broadening.}
\label{fig10}
\end{figure}

\begin{figure}
\begin{center}
\includegraphics[width=10cm]{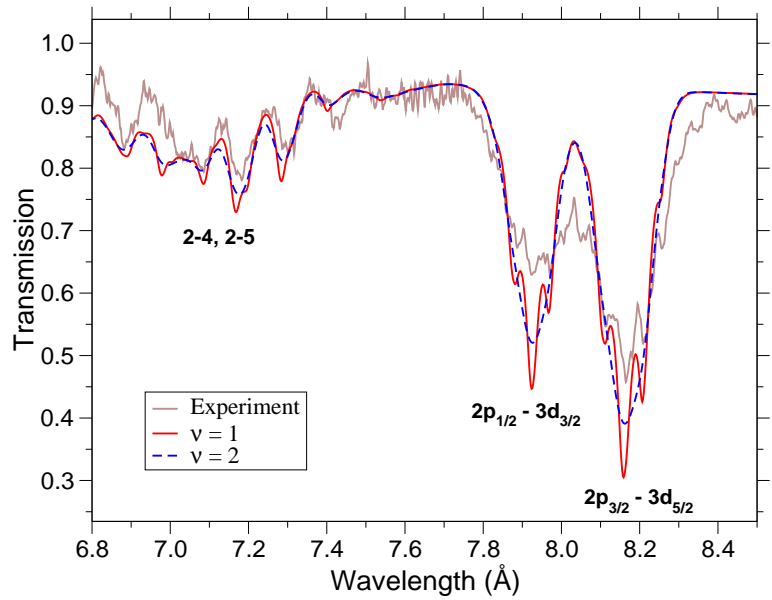}
\end{center}
\caption{(Color online) Same as Fig. \ref{fig10} for the NaBr (in the Br range)
experiment of Bailey et al. \cite{BAILEY03} at $T$=47 eV and
$\rho$=0.04 g/cm$^3$.}
\label{fig11}
\end{figure}

\begin{figure}
\begin{center}
\includegraphics[angle=-90,width=14cm]{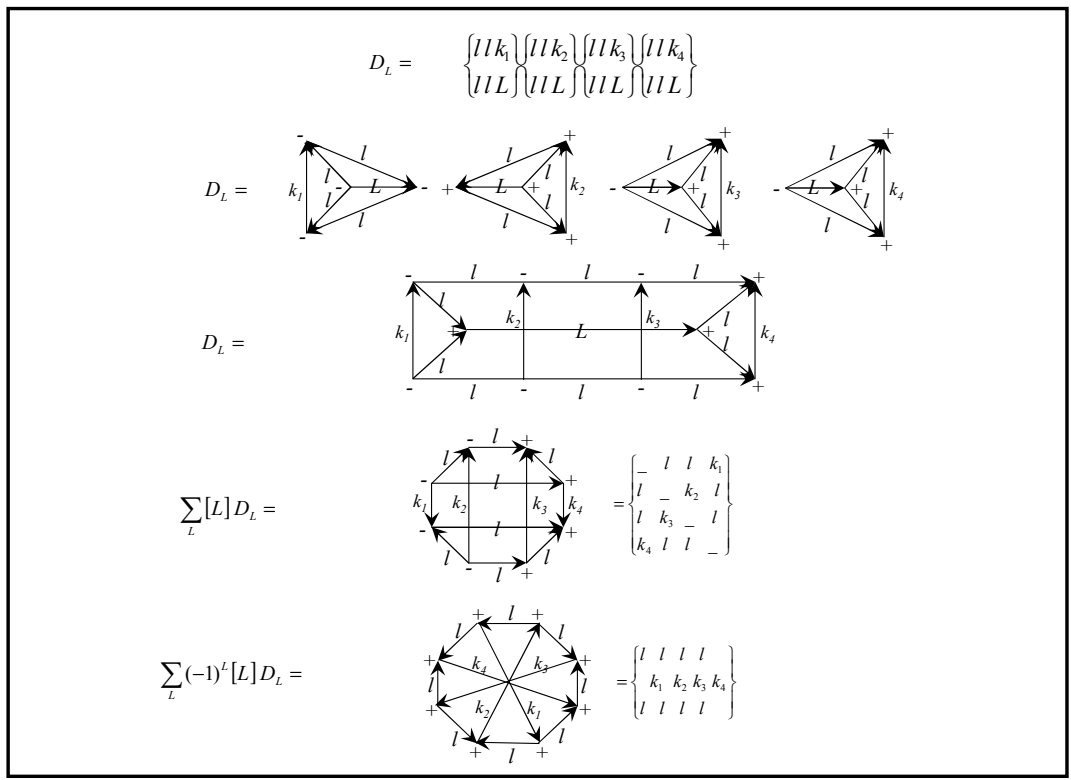}
\caption{Complexity of the calculation of a simple part of the fourth moment 
$\mu_4$ of the simple transition array $l^2\rightarrow ll'$ illustrated by
graphical methods.}
\label{fig12}
\end{center}
\end{figure}

\end{document}